\documentclass[a4paper]{article}

\usepackage{INTERSPEECH2021}
\usepackage{enumitem}
\usepackage[T1]{tipa}
\usepackage[dvipsnames]{xcolor}
\usepackage[hidelinks]{hyperref}

\title{Visualising Model Training via Vowel Space for Text-To-Speech Systems}
\name{Binu Abeysinghe, Jesin James, Catherine I. Watson, Felix Marattukalam}

\address{
  Dept. of Electrical, Computer, and Software Engineering, The University of Auckland, New Zealand}
\email{babe269@aucklanduni.ac.nz, jesin.james, c.watson, felix.marattukalam@auckland.ac.nz}

\begin{document}

\maketitle
\begin{abstract}
With the recent developments in speech synthesis via machine learning, this study explores incorporating linguistics knowledge to visualise and evaluate synthetic speech model training. If changes to the first and second formant (in turn, the vowel space) can be seen and heard in synthetic speech, this knowledge can inform speech synthesis technology developers. A speech synthesis model trained on a large General American English database was fine-tuned into a New Zealand English voice to identify if the changes in the vowel space of synthetic speech could be seen and heard. The vowel spaces at different intervals during the fine-tuning were analysed to determine if the model learned the New Zealand English vowel space. Our findings based on vowel space analysis show that we can visualise how a speech synthesis model learns the vowel space of the database it is trained on. Perception tests confirmed that humans could perceive when a speech synthesis model has learned characteristics of the speech database it is training on. Using the vowel space as an intermediary evaluation helps understand what sounds are to be added to the training database and build speech synthesis models based on linguistics knowledge.
\end{abstract}


\vspace{-2mm}
\section{Introduction}
\vspace{-2mm}
Deep learning has risen in popularity for producing high-quality synthetic speech. Common synthetic speech evaluation include subjective tests \cite{benoit1996subjective, loizou2011subjective, valizada2021sub_obj} and objective measures \cite{kominek2008objective, valizada2021sub_obj}. These evaluations are conducted \textit{after} the speech synthesis model (hereafter called speech model) is fully trained. \textit{During} the speech model training, we rely on the saturation of the learning curve to decide on when to stop the training. Even though the learning curve indicates how well the deep learning model learns the characteristics of the speech database it is trained on, it is insufficient to interpret the acoustic properties that the model learns and compare the properties to that of the database the model is trained on. Upon conducting subjective and objective evaluation, the approach is to re-train the model with a larger speech database if issues are found with the synthetic voice. This evaluation does not indicate which particular sounds in the language will need more representation in the database. 

We propose to use the  vowel space as an approach to provide an acoustic evaluation of the speech model \textit{during} training. The vowel space is a two-dimensional area bounded by the first and second formant frequency coordinates of vowels \cite{sandoval2013vowelspace}. Linguistic studies use the vowel space to study characteristics of languages \cite{jongman1989vowelspace}, accents, dialects \cite{jacewicz2007vowelspace, clopper2008vowelspace}, their changes across different regions \cite{sandoval2013vowelspace}, and time \cite{watson2018vowelspace}. New language learners are trained to produce vowels as close as possible to the language’s vowel space, and providing visual feedback using vowel spaces is often done \cite{watson2017vowelspacelearn, chao2020vowelspacelearn, dowd1998vowelspacelearn, brett2004vowelspacelearn, carey2004vowelspacelearn}. This is possible because the first and second formants can be related to the jaw opening and tongue movement, respectively. Inspired by this, \textit{we propose to use the vowel space to evaluate how a machine learning model learns the acoustics of the vowels of a language from a speech database.}  Researchers can visualise the vowel space of a given speech model and develop an understanding of how well the model learns the characteristics of the database \textit{during} training. To develop an approach to evaluate speech model training via vowel space, we answer the following questions:
\vspace{-1.5mm}
\begin{enumerate}[label=RQ\arabic*:]
    \item How does the vowel space of a speech synthesis model change during the training process? 
    \vspace{-1.5mm}
   
    \item Can humans perceive the change in vowel space of a speech synthesis model during the training process? 
\end{enumerate}
\vspace{-1mm}

\begin{figure*}[!t]
    \centering
    \vspace{-2mm}
    \includegraphics[scale=0.145]{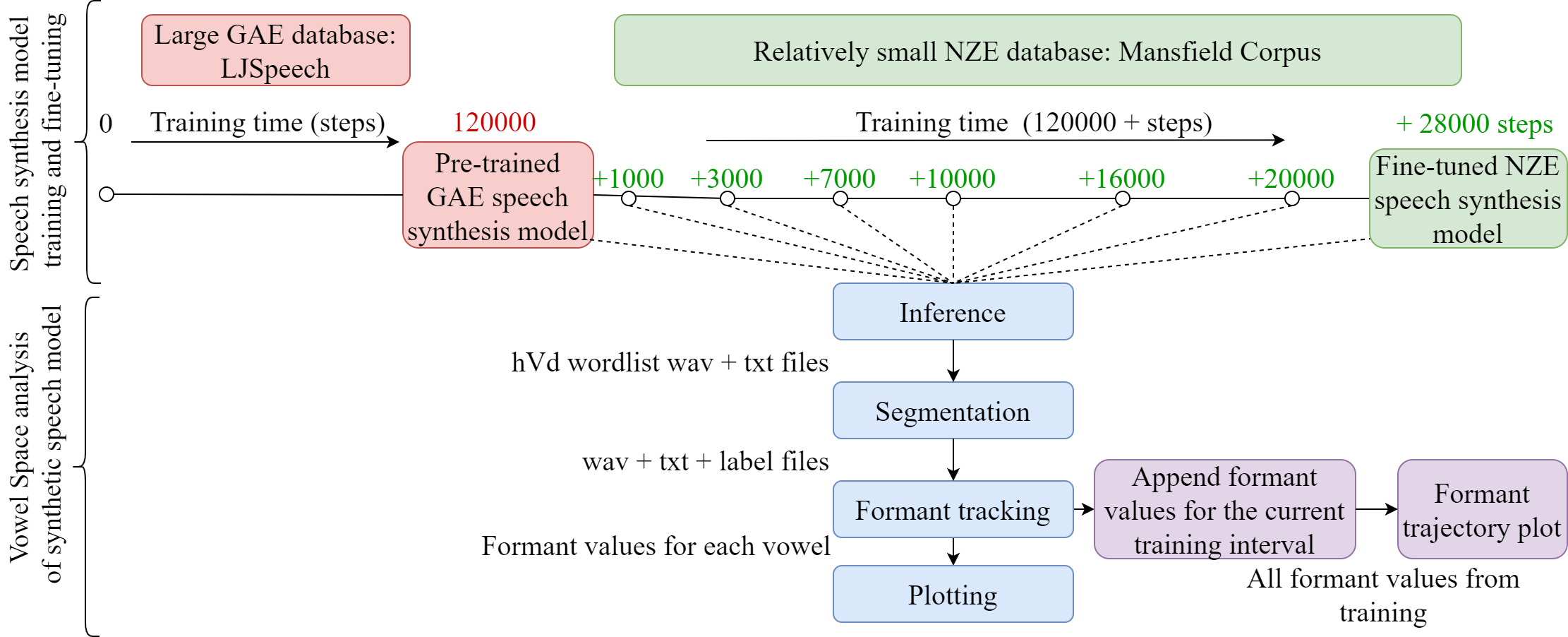} \vspace{-2mm}
    \caption{Flow diagram of the process of obtaining formant data when training a model}
    \label{fig:flow_diagram}
    \vspace{-2mm}
\end{figure*}

\begin{table*}[!t]
\centering
\resizebox{\linewidth}{!}{
\begin{tabular}{ll|lllllllllll}
\hline
1 & \color{red}NZE vowels & \color{red} \textipa{2} & \color{red} \textipa{6}* & \color{red} \textipa{I}* & \color{red} \textipa{\ae}* & \color{red} \textipa{U} & \color{red} \textipa{e}* & \color{red} \color{red} \textipa{A}:* & \color{red} \textipa{3}:* & \color{red} \textbaru:* & \color{red} \textipa{o}: & \color{red} \textipa{i}:* \\

2 & Well's lexical words & STR\color{red}{U}\color{black}{T} & L\color{red}{O}\color{black}T/CL\color{red}{O}\color{black}TH & K\color{red}{I}\color{black}T & TR\color{red}{A}\color{black}P & F\color{red}{OO}\color{black}T & DR\color{red}{E}\color{black}SS & B\color{red}{A}\color{black}TH/ST\color{red}{AR}\color{black}T & N\color{red}{UR}\color{black}SE & G\color{red}{OO}\color{black}SE & TH\color{red}{OU}\color{black}GHT & FL\color{red}{EE}\color{black}CE \\

3 & h\color{red}V\color{black}d word & h\color{red}{U}\color{black}{d} & h\color{red}{O}\color{black}d & h\color{red}{I}\color{black}d & h\color{red}{A}\color{black}d & h\color{red}{OO}\color{black}d & h\color{red}{EA}\color{black}d & h\color{red}{AR}\color{black}d & h\color{red}{EAR}\color{black}d & wh\color{red}{O'}\color{black}d & h\color{red}{OR}\color{black}de & h\color{red}{EE}\color{black}d \\
\hline
4 & Word list & \multicolumn{11}{l}{heard...hood...hud...heed...head...had...hard...hod...who'd...hood...heard...hid.} \\
\hline
\end{tabular}}
\vspace{-0.3em}
\caption{NZE vowels, usage with Well's lexical words, hVd words and example of a word list used for perception test.}
\label{tab:hvd}
   \vspace{-10mm}
\end{table*}

We train a New Zealand English (NZE) speech model to answer these questions. NZE is a distinct English accent spoken in New Zealand by 5 million people, with lexical and pronunciation differences from General American English (GAE) and British English. It is low-resourced in terms of speech technology development as there is not sufficient open-sourced data to train deep-learning-based speech models. Hence, we use a pre-trained speech model created from a large GAE database and fine-tune it using a relatively smaller NZE database. RQ1 is answered by a detailed vowel space analysis as the fine-tuning happens from GAE to NZE. RQ2 determines a link between the change in vowel space and the perception of the accent change during speech model training. A perception test and its comparison to the vowel space is conducted to answer this question.

We analyse the vowel space during speech model training, thereby providing the foundational knowledge for developing a method to visualise and evaluate speech synthesis model training. The method developed here does not replace perception tests; it provides an intermediary evaluation technique. Visualising the speech model’s vowel space allows researchers to optimise against the expected vowel space of the language/speaker. This optimisation can be done via adjustments to the model parameters or using the vowel space to determine what speech sounds need to be better represented in the training database. Once the researcher is satisfied with the speech model's vowel space, they can validate the results with a perception test.

\vspace{-2mm}
\section{Background} \label{sec:background}
\vspace{-2mm}

Formants are a frequency range where there is an absolute or relative maximum in the sound spectrum. The frequency at the maximum is the formant frequency \cite{ASA_formant_definition}. First formant, F1 corresponds to the vertical position of the tongue and is associated with frequencies between $200$ and $900$ Hz. Second formant, F2 corresponds to the horizontal position of the tongue and ranges from $600$ - $2600$ Hz \cite{watson1998acoustic} (See Figure \ref{fig:28k} (b) with the NZE vowel space marked with green line). The tongue position constricts airflow in the throat and the roof of the mouth, adjusting the frequency of the air as it passes through \cite{watsontwo}. A large part of voiced sections in speech is attributed to vowels, and vowels often exhibit a noticeable change in F1 and F2 \cite{hillenbrand1995acoustic}. There is strong evidence on the impact of vowel space on age \cite{albuquerque2019age_vowelspace}, gender and languages \cite{ tatman2017vowelspace_dialect, watson2018vowelspace}. The vowel space is a tool to evaluate articulation problems \cite{van2018vowelspace_articulation, turner1995vowelspacearea} and language learning \cite{watson2017vowelspacelearn, carey2004vowelspacelearn}, showing that it is a unique characteristic of a person’s voice for a particular language, age, gender, or speaking difficulties. 

NZE, compared to GAE, is a non-rhotic English variety, sharing many features of received pronunciation. Table \ref{tab:hvd} lists the NZE vowels using International Phonetic Alphabet (Row 1) and corresponding Well’s lexical words \cite{wells1982accents} (Row 2). The NZE vowel space is marked with the green triangle in Figure \ref{fig:28k} (b). Pertinent to this study, using Well’s lexical words, NZE has raised DRESS and TRAP vowels compared to GAE. Unlike GAE, NZE distinguishes between TRAP and BATH vowels. CLOTH vowel merges with LOT and not THOUGHT vowel in NZE. The point vowels for the NZE vowel space are FLEECE, THOUGHT, and START. The point vowels for the GAE vowel space are FLEECE, TRAP, START, and GOOSE \cite{wells1982accentsv3}.

Speech models try to learn characteristics from the database they are trained on, which in effect is learning the acoustic characteristics of the speaker(s) of the database. Although it can be argued that vowel space only captures the acoustic characteristics of the vowels, it is known that vowels are distinctive characteristics of the speaker of a language. Further, in accents of English, the main differences are due to vowel pronunciations \cite[p.181]{wells1982accents}.

\vspace{-2mm}
\section{Methodology}
\vspace{-2mm}
To answer RQ1, we analyse the vowel space during the training of a speech model from GAE to NZE as illustrated in Figure \ref{fig:flow_diagram}.

\vspace{-2mm}
\subsection{Speech databases}
\vspace{-2mm}
LJSpeech database \cite{ljspeech17} is the GAE database used, consisting $13,100$ short audio clips of one GAE female speaker, with a duration of $24$ hours. The Mansfield database \cite{watson2014resources} is the NZE database used. It consists of $1095$ sentences read by a NZE female speaker with a duration of $3$ hours.  

\vspace{-2mm}
\subsection{Speech synthesis model training and fine-tuning}
\vspace{-2mm}
A deep learning-based speech model was trained on the LJSpeech database for $120,000$ steps, as shown in Figure \ref{fig:flow_diagram}. The learning curve was observed to decide when to stop the training process. This produced the pre-trained GAE speech model. The GAE model was then fine-tuned using the Mansfield database. Again, the learning curve was observed to decide to stop the training at $28,000$ additional steps. Thus, the model was trained for $148,000$ steps with the last $28,000$ trained exclusively on the NZE database. The vowel space analysis was conducted only on the fine-tuning to observe the changes to vowel space as the model tunes from GAE to NZE.

\begin{figure*}[!t]
    \centering
    \includegraphics[scale=0.068]{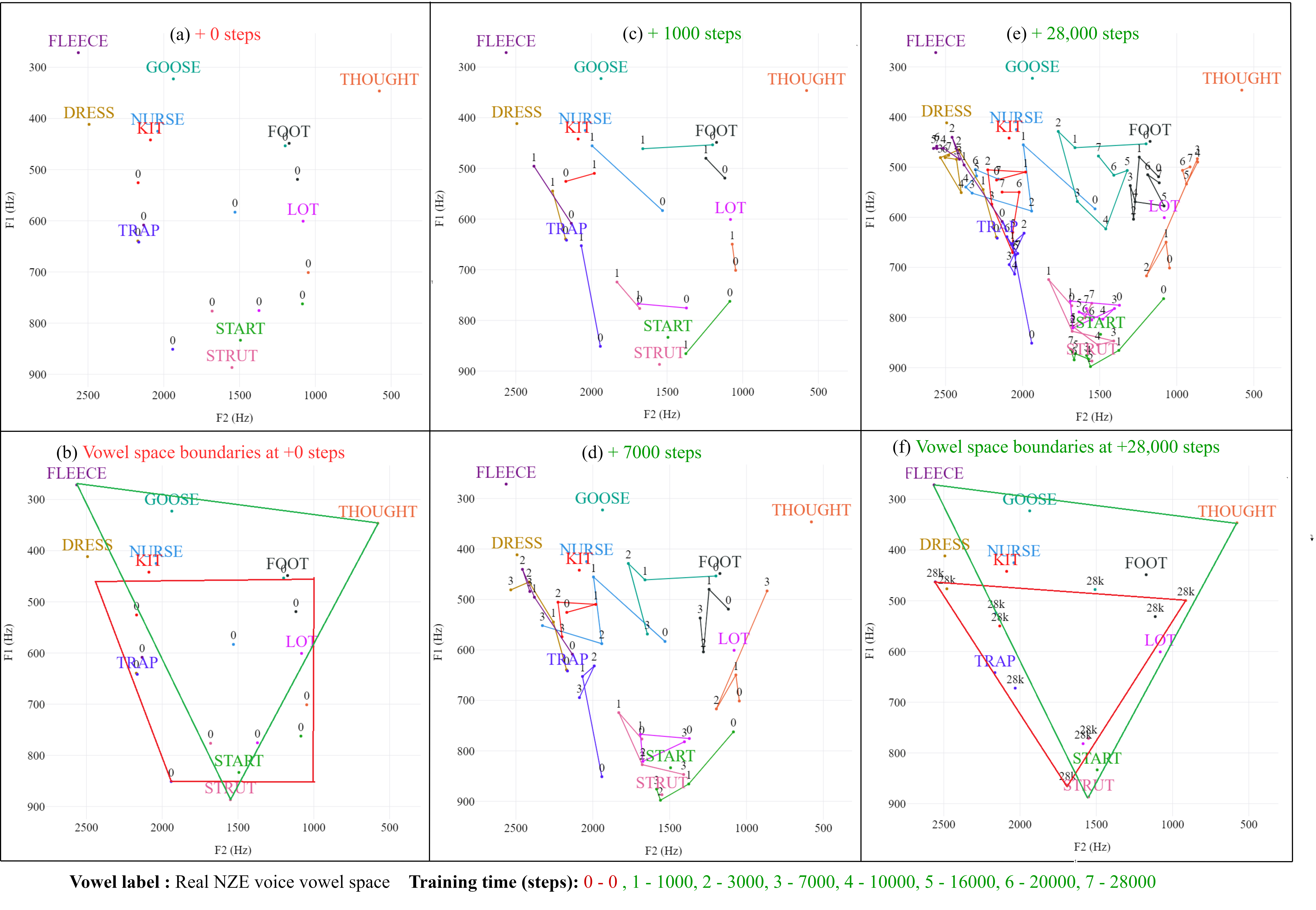}
        \vspace{-9mm}
    \caption{The change in speech model vowel space over training steps. Each vowel point is the average of multiple vowel points.}
    \label{fig:28k}
    \vspace{-7mm}
\end{figure*}

\vspace{-2mm}
\subsection{Vowel space analysis design} \label{vowel_analysis}
\vspace{-2mm}
The model training was evaluated at multiple points (all the green steps illustrated in Figure \ref{fig:flow_diagram}) during the fine-tuning. At each step, an inference of the model trained up to that step was created. This inference was used to synthesise speech corresponding to that step. Through forced alignment, the vowel segments were identified. From these, F1, F2 were extracted and used to produce a vowel space. Each vowel point is the average of multiple renderings of the same vowel (details of sentence selection in Section \ref{sentence_selection}). This vowel space represents the acoustics of vowels learned by the speech model up to that step. Figure \ref{fig:28k}(a) with points marked as $0$ is the vowel space obtained at $120,000 + 0$ steps - the fully GAE speech model. We also add the vowel space of a \textit{real} NZE voice for reference \cite{watson2014mappings} (produced by the same NZE speaker who produced the Mansfield database). Figure \ref{fig:28k}(c) with points marked as $1$ is the vowel space at $120,000 + 1000$ steps, which is the fine-tuned model after $1000$ steps. The formant values obtained at the later steps are appended along with the former steps, and lines are drawn between them to indicate the vowel space progression. In Figure \ref{fig:28k}(c) the line between points marked as $0$ and $1$ of the same color indicate the change in vowel space as the model trained from $120,000 + 0$ to $120,000 + 1000$ steps. Each colour indicates each vowel in NZE, with the \textit{real} location of the vowel represented by the Well’s lexical word of the same colour. This was then continued for the entire learning process up to $120,000 + 28000$, which gives Figures \ref{fig:28k} (d), (e).

\vspace{-3mm}
\subsubsection{Speech synthesis model and step intervals} \label{step_intervals}
\vspace{-2mm}
The auto-regressive architecture Tacotron2 \cite{Shen_Wavenet_2018} was used to train the speech model. Tacotron2 combines two deep neural networks: Tacotron’s sequence-to-sequence spectrogram generation \cite{Wang2017}, and Google’s WaveNet \cite{van2016wavenet}. Tacotron uses a Long Short Term Memory network to predict a mel spectrogram from character embeddings of the input text. The use of Tacotron in a multi-speaker scenario \cite{gibiansky2017deep} and fine-tuning Tacotron2-based pre-trained models to that of speakers or languages with limited data \cite{debnath2020low, suaracu2021low} prove that smaller datasets could benefit from transfer learning using Tacotron2. Initially, the model was trained using the LJSpeech database for $120,000$ steps based on \cite{tensorflowtts}. The $120,000$ steps pre-trained model was then trained for a further $28,000$ steps using the Mansfield database. The fine-tuning was conducted with an initial learning rate of $0.001$ on Google Colab \cite{google_2017}. Nvidia Tesla K80 GPU was used for training, taking approximately 48 hours. 

 Conducting vowel space analysis at every step is time-consuming. The authors found that the most perceptual changes to the synthetic voice from GAE to NZE happened at the first $10,000$ steps compared to the last $10,000$ steps. Hence, more steps were chosen earlier during the fine-tuning. The steps chosen were: $0$, $1000$, $3000$, $7000$, $10,000$, $16,000$, $20,000$, $28,000$ steps; but analysis can also be conducted at all steps.

\vspace{-3mm}
\subsubsection{Sentence selection} \label{sentence_selection}
\vspace{-2mm}
Following linguistics studies \cite{watson2014mappings}, hVd words were chosen to avoid  co-articulation effects (Table \ref{tab:hvd} Row $3$). Five word lists were produced with the 11 hVD words in different orders to ensure that the variations in synthetic speech produced based on the neighbouring words and position of the word in a sentence were balanced out as described in \cite{watson2014mappings}. Ellipses were added between each hVd word (Table \ref{tab:hvd} Row $4$) to introduce a pause between them during inference. As we are fine-tuning to NZE, the analysis aims to identify how well the speech model learns NZE. Hence, the analysis considered only the NZE vowels.

\vspace{-3mm}
\subsection{Vowel space analysis method}
\vspace{-2mm}
Figure \ref{fig:28k} shows the vowel space analysis steps. Word lists were given to the inference of fine-tuned $0$ step NZE model to synthesise corresponding wav files. wav files were segmented at the phonetic level using WebMAUS \cite{schiel1999}, which has an NZE option. Text, wav file and its phonetic transcription were used to extract formants using $formant\textunderscore burg()$ in Praat \cite{praat} with default settings. The averages of formant values for each vowel were used to plot a vowel space using Python’s $plotly$ library \cite{plotly}. This was repeated for selected steps to obtain Figures \ref{fig:28k}(a),(c),(d),(e), taking 40 s per step, mainly due to using external segmentation service. Based on the resources available, the analysis times can be optimised. Hand correction of segmentation and formant estimation was not done to facilitate automation. 

\begin{figure*}[!t]
    \centering
    \includegraphics[scale=0.15]{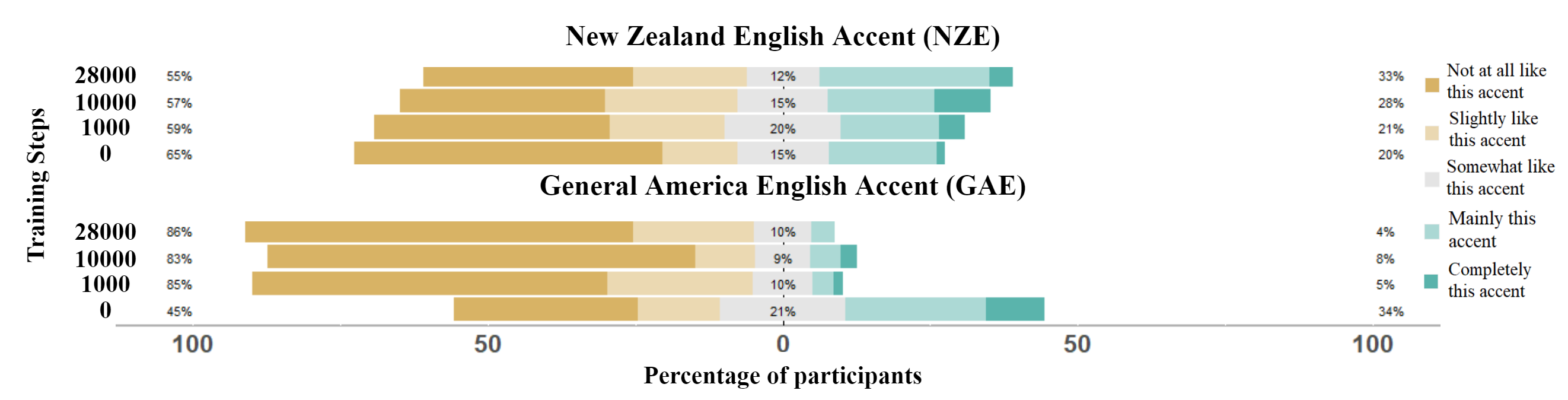}
        \vspace{-3mm}
    \caption{Perception test results in a Likert scale represented using stacked bar graphs. The accent given at the top of each stacked bar plot is the accent option that was provided to participants to choose from.}
      \vspace{-7mm}
    \label{fig:likert}
\end{figure*}

\vspace{-2mm}
\subsection{Perception test} \label{sec:perception_test}
\vspace{-2mm}

A perception test\footnote{Approved by the University of Auckland Human Participants Ethics Committee (Ref.22681) on 01/10/2021 for 3 years.} was conducted to understand if the changes in what is learnt by the speech model \textit{during}  training are perceivable, thereby addressing RQ2. Sentences with words containing the NZE vowels along with a carrier sentence were used. The carrier sentence used was `Say the word ... again’, where a word replaced the ellipses. Based on Section \ref{sec:background}, the vowels marked by * in Table \ref{tab:hvd} are the most different between GAE and NZE. Two words containing each vowel were chosen and placed into the carrier sentence. This resulted in $16$ sentences (two each for the $8$ vowels marked by *) at each of the $8$ step intervals, giving a total of $128$ sentences.  An online perception test was conducted where the participants listened to 42 synthetic speech sentences. They rated the sentences using a Likert scale ranging from \textit{`Not at all like this accent’} to \textit{`Completely this accent’}. Along with GAE and NZE, Australian English (being similar to NZE \cite{watson1998acoustic}) and Canadian English (being similar to GAE \cite{labov2008atlas}) were included as options. Depending on the perceived closeness to an accent, participants could move a slider on one or all the accents. 23 people participated in the survey, aged 22-57 (mean = 33, SD = 12). 13 were first language English speakers. All participants had lived in New Zealand for more than a year and considered themselves proficient in English. None of the participants reported any hearing difficulties.

\vspace{-2mm}
\section{Results and discussion}
\vspace{-2mm}

\subsection{Vowel space analysis results}
\vspace{-2mm}
The vowel space analysis resulted in vowel spaces for selected steps as shown in Figure \ref{fig:28k} (a), (c), (d), (e). It can be seen that there is a change in the vowel space during fine-tuning from one accent to another. The diagram illustrates that the process of vowel space transformation during speech model training is far from a linear progression. Additionally, they can be used to understand if the model can learn the vowel space accurately. We have also included the boundaries for the vowel spaces learned at $0$ steps (Figure \ref{fig:28k} (b)) and at $+28,000$ steps (Figure \ref{fig:28k} (e)). Contrasting the red vowel spaces in Figure \ref{fig:28k} (b) and (f), one can see how it changes from the GAE trapezoid shape to the NZE triangular shape, with the point vowels FLEECE, THOUGHT and START, similar to the NZE vowel space derived from the real recordings, given in green. The difference in vowel height between the two NZE vowel spaces (Figure \ref{fig:28k} (f) is most likely due to differences in speaking style. The real speech was spoken carefully, with hyper-articulation. The synthetic speech was produced with little pause between the words, mimicking the continuous speech it is trained on.

The figures also show how well the model can replicate the vowel space of the database it is trained on. All vowels change the most rapidly within the first $10,000$ steps (Figures \ref{fig:28k} (c), (d), (e)). Smaller variation is noticed as the model trains above $20,000$ steps (Figure \ref{fig:28k}(e) compared to (d)). This is most evident in the TRAP, LOT, STRUT, FOOT vowels. These results show that the vowel space of the speech model changes towards the vowel space of the database that it is trained on, thereby answering RQ1.

\vspace{-2mm}
\subsection{Perception test results}
\vspace{-2mm}

The summary of the results for each accent option at four step intervals is shown in Figure \ref{fig:likert}\footnote{Results of all selected steps and the other two accent options are not included in the figure due to space constraints. For the most part, Canadian English and Australian English behaved similar to GAE and NZE, respectively.}. The shades of green indicate that the participants thought that the given synthetic sentence was \textit{mainly/completely }the listed accent. At $0$ steps $34\%$, participants thought that the speech samples were \textit{mainly/completely} GAE. As the steps increase, the number of people who think that the speech sample given was \textit{mainly/completely }GAE reduces (green portion decreases to $4\%$ at $28,000$ steps). Looking at the NZE result, as the number of steps increases, the number of people who perceived the speech samples as NZE also increases ($20\%$ in step $0$ to $33\%$ in step $28,000$). After around $10,000$ steps, the increase in identification seems to saturate. It is also important to note the large number of zero values present in each category indicates that participants generally associated the samples as one accent rather than a mix of one or more accents. Observing shades of brown at $0$ steps, $45\%$ participants thought that the GAE voice \textit{did not sound like/slightly sounded like} GAE, and $65\%$ of the participants thought that the NZE voice \textit{did not sound like/slightly sounded like} NZE. 

The perception test results can be linked to the vowel space analysis. Using the vowel space analysis (Figure \ref{fig:28k}) we can see how the vowel space changes as the training progresses. From the perception test, it can be seen that the participants also perceived a change in accent from GAE to NZE as the step numbers increases. We can see that the major change in the vowel space from GAE to NZE happens in the first $10,000$ steps of the training. A similar trend was observed in the perception test, where there was a sudden drop in the number of participants who perceived the synthetic speech samples as \textit{completely/most likely} GAE at $1000$ steps and an increase in participants who perceived the synthetic speech samples as\textit{ completely/most likely} NZE. Hence, RQ2 has been answered. However, the test sentence `Say the word ... again’ may have been insufficient to highlight the differences between vowels. More phonetically varied sentences could be chosen for detailed insights into the impact on each vowel. The process of obtaining these plots have been automated into a Python package dubbed \textit{performant}\footnote{\url{https://github.com/babe269/performant}}.

When deep learning models give inaccurate results, the solution is to train with more data or adjust training parameters. The vowel space analysis can help researchers understand which vowels need to be incorporated into the database. Also, the vowel space shape can be included as a loss function to be optimised during training. As yet, a measure for changing vowel space shape has not been automated. Also, with transformer models being used for speech synthesis, the specific vowels crucial for a language can be given more attention. Finally, the changing vowel space (Figure \ref{fig:28k}) can be used by researchers as an indicator to decide when to stop training a deep learning model, along with the learning curve which is usually used.

\vspace{-2mm}
\section{Conclusion}\vspace{-2mm}
This study sought to explore a heavily researched aspect of linguistics - vowel spaces, and see if they were as helpful in evaluating deep learning-based synthetic speech models. This study provides researchers an easy means to visualise the closeness of a speech model to what is expected for a language, accent or speaker \textit{during} the model training. It provides researchers with an alternative approach for evaluating deep learning models and helps better visualise the `learning’ of these complex models. Vowel space also paves the way to develop deep learning approaches that can adjust the model’s training to learn the vowel space accurately. This study was conducted in the hope that future speech research will consider vowel space accuracy more closely when developing text to speech models. The work conducted in this study will give rise to better quality synthetic voices grounded on linguistics knowledge.

\vspace{-2mm}
\section{Acknowledgements}
\vspace{-2mm}
The authors thank the participants of the perceptions test and the University of Auckland Faculty of Engineering for funding.

\bibliographystyle{IEEEtran}

\bibliography{mybib}

\end{document}